\documentclass[twocolumn]{aastex7}
\submitjournal{RNAAS}
\begin{document}

\title{Limits to star formation in post starburst galaxies from Type II supernovae}

\author[]{Laetile Makoko} 
\affiliation{Botswana International University of Science and Technology}
\email{}

\author[orcid=0000-0003-1455-7339,sname='North America']{Roberto De Propris}
\affiliation{Finca, University of Turku}
\altaffiliation{Botswana International University of Science and Technology}
\email[show]{depropris.roberto@gmail.com}

\begin{abstract}

We establish significant upper limits to the current star 
formation rate in two samples of post starburst galaxies 
by measuring the rate of Type II supernovae from the Zwicky
Transient Factory Bright Transient Survey. No Type II 
supernovae are observed within the Petrosian radii of $z < 0.05$ post starburst galaxies during this supernova search survey. We calculate that at 95\% confidence level the star formation rate in these galaxies is $< 0.8$ M$_{\odot}$ yr$^{-1}$.

\end{abstract}



\section{Introduction} 

Post-starburst galaxies (hereafter PSGs) are rare objects that are located in the ‘green valley’ of the galaxy colour–magnitude diagram \citep{Wong2012}. Their optical spectra are dominated by A-type stars, suggesting that a recent burst of star formation was terminated abruptly $\lesssim 1$ Gyr ago. However, PSGs show no nebular emission consistent with on-going star formation activity. PSGs may offer clues as to how galaxies initiate and quench star formation. A significant fraction of massive galaxies at $z < 1$ must have undergone a post starburst phase \citep{Wild2020}. 

The lack of strong emission lines and the observed colours tend to exclude any current star formation. A strong limit to current star formation activity in PSGs is provided by the 1.4 GHz flux of stacked sources: \cite{Nielsen2012} find that the average star formation rate in PSGs must be $< 1.6 M_{\odot}$ yr$^{-1}$, with most of the radio flux coming from a small fraction of objects and likely weak AGNs. However, PSGs often contain substantial gas reservoirs potentially able to fuel star formation (e.g., \citealt{Li2024,Ellison2025}). Significant molecular gas reservoirs have been found in optically selected PSGs (e.g., \citealt{Yesuf2020,Bezanson2022,Smercina2022} and references therein). Some PSGs may also have active star formation in heavily obscured regions \citep{Poggianti2000,Baron2023}.

Here we estimate the current star formation in PSGs using the
Type II supernova rate. Type II supernovae come from massive
($> 8 M_{\odot}$) stars and therefore yield an estimate of 
the recent ($\sim 50$ Myr) star formation rate. In the next section we detail the datasets we use and our analysis. Discussion and conclusions are presented in Section 3. We use the cosmological parameters from \cite{Planck2020} where needed.
%% The "ht!" tells LaTeX to put the figure "here" first, at the "top" next
%% and to override the normal way of calculating a float position.
%% The asterisk after "figure" tells the compiler to span multiple columns
%% if a two column style is selected.

\section{Data}

There are two large datasets of nearby PSGs: the one by \cite{Goto2007} as expanded in \cite{Nielsen2012} and 
analysed by \cite{Melnick2013} and the `shocked' sample
of \cite{Alatalo2016} where some of the constraints on
line emission are relaxed to select younger objects.

We search for Type II supernovae within the Petrosian radius
of each galaxy (taken from the SDSS: \citealt{York2000,Abdurrouf2022}) from the Zwicky Transient Facility Bright Transient Survey \citep{Fremling2020}. This survey classifies all transients spectroscopically and is stated to be complete for Type II supernovae to at least $z=0.05$ (\citealt{Fremling2020}, see their Figure 4, also see \citealt{Irani2022}).

\section{Results and Discussion}

From the PSG sample of \cite{Melnick2013} we find 0 supernovae in 65 galaxies, while from the sample of \cite{Alatalo2016} we find 0 supernovae in 246 galaxies.

We use the binomial distribution to estimate upper limits to the fraction of supernovae in these galaxies. At the 95\% confidence level the upper limit to the supernova fraction is
0.04 for the \cite{Melnick2013} sample and 0.01 for the \cite{Alatalo2016} sample. The two samples have some overlap but they have a somewhat different selection, with \cite{Alatalo2016} supposed to select for younger objects and we therefore prefer to treat them separately.

We can now use the expressions from \cite{Ma2025} to calculate the volumetric supernova and star formation rates for these samples.

$$SNR_{CC} = k \times SFH$$

where SFH is the star formation history as a function of redshift (in this case we have a single redshift range) and,

$$ k = {\int_{8 M_{\odot}}^{50 M_{\odot}} \Psi(M) dM \over 
{\int_{0.1 M_{\odot}}^{125 M_{\odot}} M \Psi(M)dm}} = 0.0070 M_{\odot}^{-1}$$

where $\Psi(M)$ is the Salpeter mass function 

Hence:\\

$SFH$ (\citealt{Melnick2013}) $< 0.06$ $M_{\odot}$ yr$^{-1}$ Mpc$^{-3}$\\

$SFH$ (\citealt{Alatalo2016}) $< 0.02$ $M_{\odot}$ yr$^{-1}$ Mpc$^{-3}$\\

This is equivalent to a SFR of $< 0.8$ and $< 0.3$ $M_{\odot}$ yr$^{-1}$ respectively. which is in good agreement with the upper limit derived from 1.4 GHz observations in \cite{Nielsen2012}

We can use the expression in \cite{Ma2025} to derive the volumetric supernova rate:

$${\rm SNuM(M_0)}={N \over T} {M_0^{RSS1} \over \Sigma M_i^{RSS_M+1}}$$

where $N$ is our upper limit, $T$ is the duration of the BTS survey (7 years), $M_0$ is a reference mass of $4 \times 10^{10} M_{\odot}$ and the sum is carried out over the masses of all galaxies we consider: $RSS_M$ is the rate-size slope derived by \cite{Li2011}. We adopt $RSS_M=-0.25$. 

$${\rm SNuM(M_0)} < 0.008\ {\rm Melnick\ \&\ De Propris\ 2013} $$

$${\rm SNuM(M_0)} = < 0.0004\ {\rm Alatalo\ et\ al.\ 2016}$$

in units of SN(100 yr$^{-1}$ {(10$^{10}$ M$_{\odot}$)$^{-1}$})

\begin{figure}
\vspace{0.5cm}
\includegraphics[width=\linewidth]{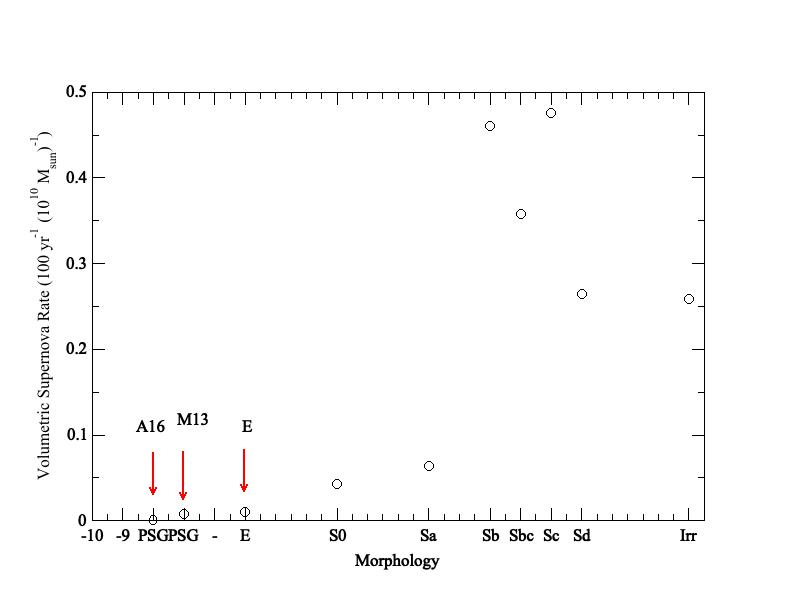}
\caption{Volumetric supernova rates from \cite{Ma2025} and our results vs. Hubble type. Upper limits are indicated by the downward facing arrows and identified in the figure (to the left of each arrow). Error bars from \cite{Ma2025} are omitted.}
\label{fig:fig1}
\end{figure}

In Fig.~\ref{fig:fig1} we compare these upper limits with
the volumetric Type II supernova rates for galaxies of several Hubble types within the GLADE+ 40 Mpc sample of galaxies in \cite{Ma2025}. 

Our upper limits are comparable to the upper limits for ellipticals reported in \cite{Ma2025}. We therefore conclude that star formation in these objects has completely ceased. All residual radio signal is likely due to weak AGNs that may not suffice to suppress star formation.

\begin{contribution}
%%This section gives authors the space to recognize author contributions. The text inside this environment is NOT counted towards the total word quanta. At a minimum, manuscripts are expected to include this text:

All authors contributed equally 
%% But authors are expected to provide more specific details, e.g. 
%%
%%SC was responsible for writing and submitting the manuscript.
%%WWM came up with the initial research concept and edited the manuscript.
%%OTS obtained the funding and edited the manuscript.
%%EBF provided the formal analysis and validation. He also edited the manuscript.
%%GEH Supervised the undergraduates, wrote the software and administers the project github and Zenodo repositories.
%%
%% Authors can use the Contributor Role Taxonomy (CRediT) at
%% for ideas on how write a good statement tailored to their needs.

\end{contribution}

\facilities{SDSS, ZTF}

\software{R, Topcat \citep{Taylor2011}
          }


\bibliography{sample7}{}
\bibliographystyle{aasjournalv7}



\end{document}